\documentclass[12pt]{article}
\usepackage{latexsym,amsmath,amssymb,amsbsy,graphicx}
\usepackage[utf8]{inputenc}
\usepackage[T2A]{fontenc}
\usepackage[space]{cite} % If exist.
\usepackage{multirow}

\usepackage{graphicx}% Include figure files
\usepackage{dcolumn}% Align table columns on decimal point
\usepackage{bm}% bold math
\usepackage{todonotes}

\makeatletter\def\@biblabel#1{\hfill#1.}\makeatother

\begin{document}

\noindent\begin{minipage}{\textwidth}
\begin{center}

{\large Evidence for symbiotic nature of 2MASS J21012803+4555377}

A.M.~Tatarnikov, A.A.~Tatarnikova, N.A.~Maslennikova, A.V.~Dodin, M.A.~Burlak, A.A.~Tatarnikov\\[6pt]

\textit{Lomonosov Moscow State University, Faculty of Physics, Moscow, 119991 Russia}
\textit {E-mail: $^a$andrew@sai.msu.ru} \\ [1cc] 
\end{center}

{\parindent5mm Just less than 300 symbiotic stars are currently known in the Galaxy. The population synthesis methods predict that this amount should be 10--100 times larger. In recent years, several works have attempted to find symbiotic candidates from photometric surveys. Regular spectroscopic observations of these candidates can increase the number of known symbiotic systems. We aim to verify the symbiotic nature of 2MASS J21012803+4555377. Archive photometry and low-resolution spectra ($\lambda/\Delta\lambda\sim1500-2000$) of the candidate are presented. Besides molecular bands, the spectrum displays numerous emission lines including forbidden and high-ionization ones -- He\,II, [Fe\,VII], etc. A strong Raman-scattered O\,VI line is observed at $\lambda6825$. The source shows prominent excessive emission in the infrared related to a dense circumstellar dust shell. All this evidence indicates that 2MASS J21012803+4555377 should be classified as a D-type symbiotic star.
\vspace{2pt}\par}

\vspace{1pt}\par

\vspace{1pt}\par
\end{minipage}

\section*{Introduction}

Symbiotic stars (SySts) are interacting binaries consisting of a red giant and a hot component. In most systems, the hot component is a white dwarf or a hot subdwarf. There are nearly 300 SySts in the Galaxy known so far \cite{Catalog2019}, \cite{Merc_2019}, and it is much less than theoretical predictions. Thus, \cite{Lu2006} estimated the number of SySts with white dwarf accretors as 1200-15000, whereas according to \cite{Magrini2003}, the total amount may be greater than $4 \cdot 10^5$. The search for new SySts is hindered by the necessity to repeat spectroscopic observations, since the emission spectrum of SySts in the passive state may be very weak. 

In recent years, some works have been performed to find the SySt candidates from different photometric surveys using machine learning, e.g., \cite{Akras_2019}, \cite{Rimoldini_2023}. \cite{Rimoldini_2023} were the first to classify 2MASS J21012803+4555377 as a SySt candidate basing on photometric and astrometric Gaia DR3 data. Before 2023, the star was considered a long-period variable with a period of nearly 486~d (\cite{Heinze_2018}, \cite{Lebzelter_2023}).

We aim to verify the symbiotic nature of 2MASS J21012803+4555377 and estimate the parameters of the system.

\section*{Observations}

Low-resolution spectroscopic observations were carried out on Jan 11, 2025 (and repeated on Jan 15) with the Transient Double-beam Spectrograph (TDS, \cite{Potanin2020}) mounted on the 2.5-m telescope of the Caucasian Mountain Observatory of the Sternberg Astronomical Institute of the Lomonosov Moscow State University (CMO, \cite{Shatsky2020}). A $1''$ slit was used, the blue and red spectral regions were obtained simultaneously through the $B$ (spectral range $360-577$~nm, spectral resolution $R=1300$) and $R$ (spectral range $567-746$~nm, spectral resolution $R=2500$) arms of the instrument. The primary data reduction was conducted using the pipeline described in \cite{Potanin2020}.

We also carried out observations with the new fiber-fed echelle spectrograph prototype mounted on the 2.5-m telescope of CMO on Jan 11, 2025. The spectrograph has a 2.5~arcsec round aperture and produces spectra between 4100-8200{\AA} with a resolving power of $R \approx 20000$, sampled with 0.042\AA/pix in 31 spectral orders. Unfortunately, the short exposure time allowed us to detect only the H$\alpha$ line in the spectrum.

Photometric observations were carried out on Jan 12, 2025 on the 60-cm telescope (RC600, \cite{Berdnikov_2020}) of CMO with FoV $\approx20'$ and $0.7''$/pix scale. We obtained the $VR_cI_c$ magnitudes by means of differential aperture photometry using field stars with magnitudes $V\sim13-14$~mag taken from the APASS catalogue \cite{Henden_2016}. We derived $V=16.635$, $R_c=14.056$, $I_c=12.25$ for 2MASS J21012803+4555377.   

\section*{Results of observations and discussion}

\begin{figure}
\includegraphics[width=\hsize]{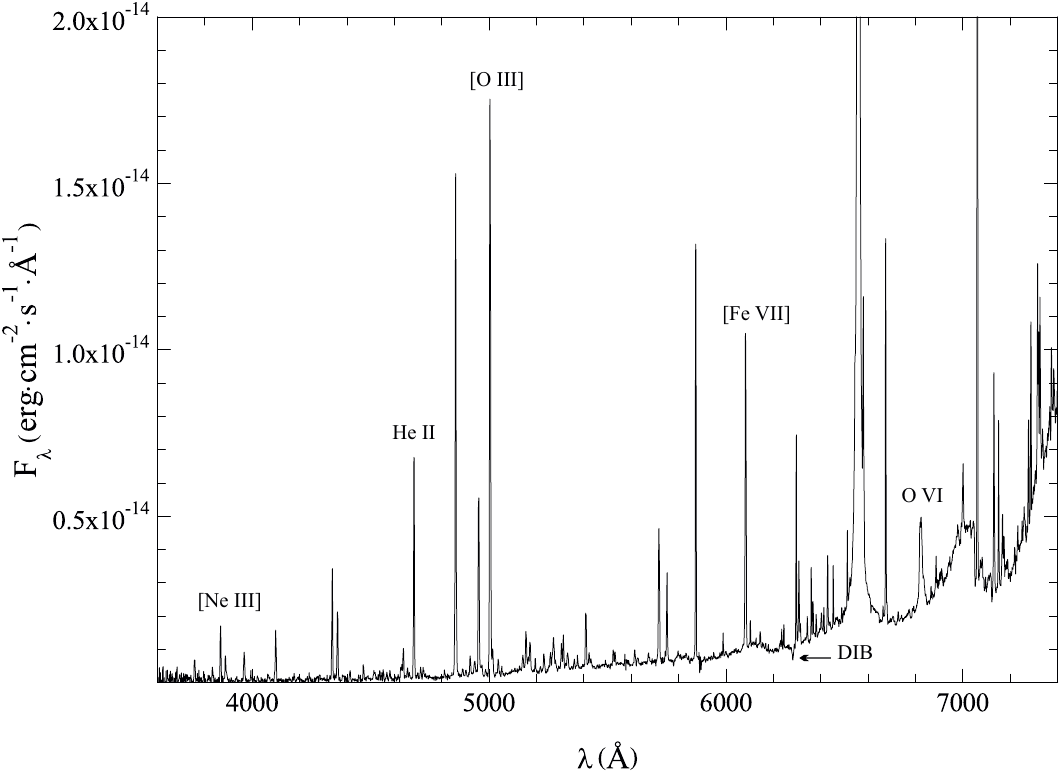} 
\caption{The spectrum of 2MASS J21012803+4555377 obtained on Jan 11, 2025 (the spectrum was flux calibrated using $V=16.635$).} 
\label{fig:spectr_obs}
\end{figure}

In Fig.~\ref{fig:spectr_obs} we present the spectrum of 2MASS J21012803+4555377 exhibiting typical features of a SySt. A rich emission-line spectrum including high ionization lines (He\,II, [Fe\,VII], etc.) is superimposed on the continuum of the cool component with noticeable molecular bands (TiO $\lambda7050$, etc.). There is also a wide emission feature at 6825\,\AA{}. It is known to be a Raman-scattered O\,VI line \cite{Schmid1989}. This feature has been detected so far only for SySts, and therefore, it is an important criterion which can help to distinguish them from other types of variables \cite{Catalog2019}. So, 2MASS J21012803+4555377 should be classified as a SySt.      

Besides the absorptions produced by the cool component, the spectrum shown in Fig.~\ref{fig:spectr_obs} also displays features related to interstellar extinction, for example, the prominent DIBS at $\lambda6284$ and $\lambda 5780$ and Na\,I resonance doublet. 

\begin{figure}
\centering
\includegraphics[width=1\textwidth]{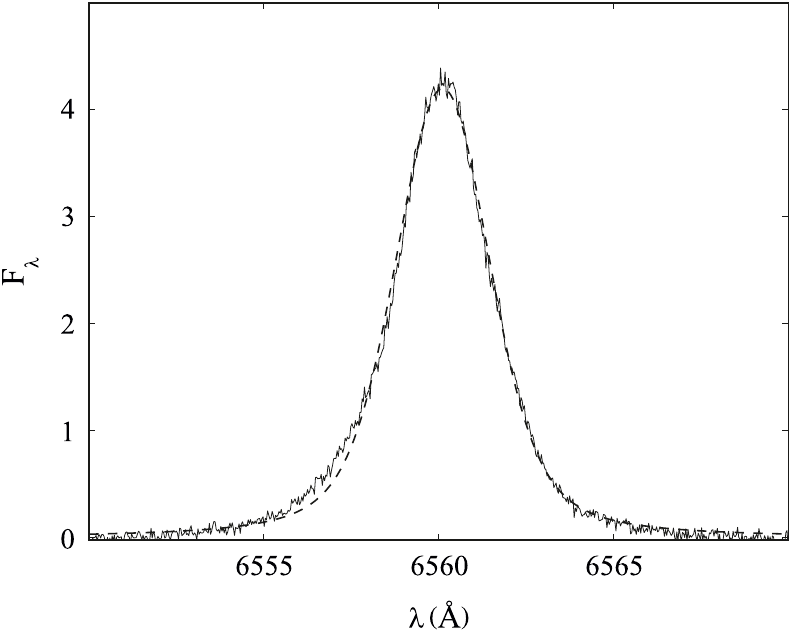} 
\caption{The profile of H$\alpha$ and the Voigt profile (dashed line) corrected to the Solar system baricenter.} 
\label{fig:Ha_hrs}
\end{figure}

We measured the radial velocity of 2MASS J21012803+4555377 by fitting the H$\alpha$ line in high-resolution spectra with a one-component Voigt profile (fig.~\ref{fig:Ha_hrs}): $V_r=-121.7\pm0.5$~km/s and FWHM~$=3.1$~\AA{}.

\begin{figure}
\centering
\includegraphics[width=0.6\textwidth]{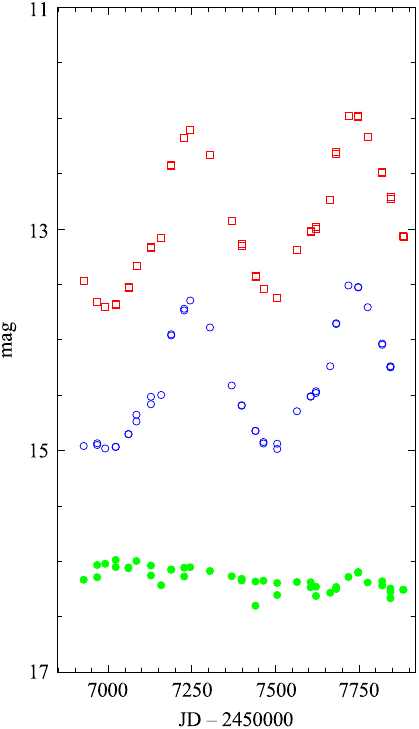} 
\caption{The light curves of 2MASS J21012803+4555377 in the $Bp$ (green), $G$ (blue) and $Rp$ (red) bands obtained by Gaia in 2015~-- 2017.} 
\label{fig:light_curves}
\end{figure}

Fig.~\ref{fig:light_curves} displays \textit{Bp, G} and \textit{Rp} light curves obtained by Gaia in 2015-2017 \cite{Gaia_DR3}. One can see that there are no noticeable variations in the $Bp$ band, whereas the $G$ and $Rp$ curves demonstrate well-expressed brightness oscillations with a peak-to-peak amplitude of $\sim$1.5~mag and the time between maxima of nearly $488^{\rm d}$. There is a hump on the rising branch -- a deceleration in brightening. Such feature appears in phase curves of some Mira variables (see, e.g., \cite{Lockwood1971}, \cite{Nadzhip2001} and \cite{Fedoteva2020}). If we consider $488^{\rm d}$ to be the pulsation period of the cool component, then it falls in the middle of the range of periods typical for Mira stars. But the amplitude of visual brightness variation has to be not less than 2.5~mag for Mira variables \cite{Samus2017} and  decreases with wavelength \cite{Lockwood1971}. That 2MASS J21012803+4555377 does not show brightness oscillations in the visual range must be due to the dominating emission from the nebula (and perhaps hot component) which veils the TiO bands (see Fig.~\ref{fig:spectr_obs}). The input from the nebula is less in the red part of the spectrum, therefore we do see the brightness variation of the cool component, although with a reduced amplitude.

\begin{figure}
\includegraphics[width=1\textwidth]{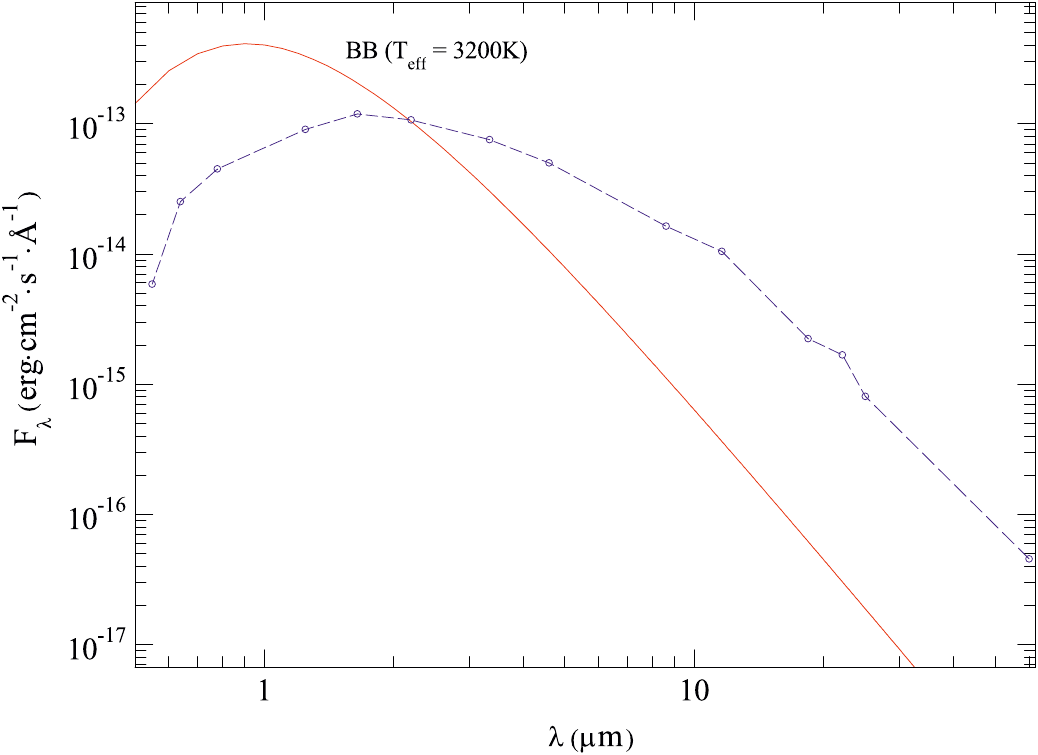} 
\caption{The plot showing the SED of 2MASS J21012803+4555377 corrected for interstellar extinction with $E(B-V)=0.7$~-- blue line and the emission of a $T_{\rm eff} = 3200$K blackbody with the same bolometric flux~-- red line. The circles on the blue line represent our $VR_cI_c$ photometry, the $JHK_s$ data from 2MASS \cite{Skrutskie2006}, and also WISE \cite{WISE}, AKARI \cite{AKARI}, and IRAS \cite{IRAS} data.}
\label{fig:sed}
\end{figure}

According to \cite{Bailer-Jones2021}, the median photogeometric distance of 2MASS J21012803+4555377 is 2812~pc (the 16\% quantiles of a posterior probability distribution are 2312 and 3382~pc). \cite{Green2019} gives $E(B-V)\approx0.7$ for this distance in the direction of the source, but reducing distance to 2500~pc results in a much smaller value of $E(B-V)\approx0.3$.

In Fig.~\ref{fig:sed} we present the spectral energy distribution (SED) of 2MASS J21012803+4555377 corrected for interstellar extinction with $E(B-V)=0.7$, using the extinction curve of \cite{Fitzpatrick1999}. One can see that emission peaks at $\lambda\sim2$~$\mu$m. The bolometric flux is $F_{bol}=5.7\cdot10^{-9}$~erg/(cm$^2$\,s\,\AA{}). As the major part of energy is emitted in the infrared, $F_{bol}$ barely depends on $E(B-V)$, and for the adopted distance of 2.8~kpc the components of the system inside dust shell will have a luminosity of 2130~L$_\odot$. This value is typical for SySts.     

\begin{figure}
\includegraphics[width=1\textwidth]{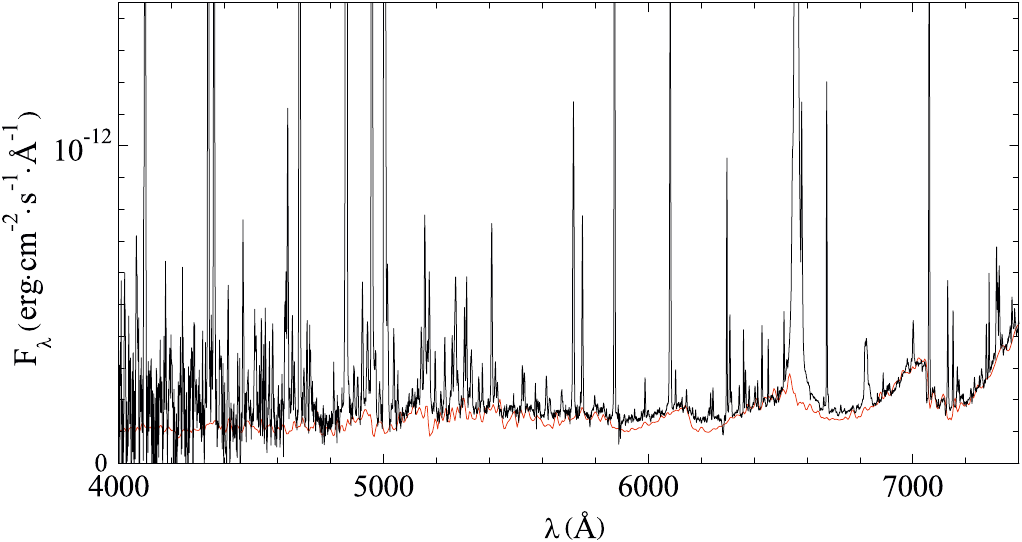} 
\caption{The low-resolution spectrum of 2MASS J21012803+4555377 corrected for interstellar extinction and extinction in the circumstellar shell with the total $E(B-V)=2$ (black line) and the sum of continua of an M5\,III red giant and a nebula with $T_e=15000$K (red line).}     
\label{fig:spectr_M5}
\end{figure}

The SED displays a significant IR excess emission related to a dense circumstellar shell. The extinction in the shell may be estimated by the ratio of fluxes at wavelengths where we can neglect the emission from dust (Fig.~\ref{fig:sed}). If we assume that the wavelength dependency of the circumstellar extinction is similar to the interstellar extinction law, we derive $E(B-V)\approx1.3$ for the extinction in the shell. So, the total extinction is $\approx2$~mag. 

The presence of strong He\,II $\lambda4686$ line implies a high temperature of the hot component. If we assume that the continuum around the line is emitted completely by the nebula and hot component, then the equivalent width of the line $\rm {EW(He\,II)}=250$\,\AA{} gives a temperature of $T_{hot}\approx110000K$ for the hot component (case B recombination). This may be a low estimate of $T_{hot}$.  

Fig.~\ref{fig:spectr_M5} shows the spectrum of 2MASS J21012803+4555377 corrected for extinction with $E(B-V)=2$ following the equation from \cite{Fitzpatrick1999}. The TiO molecular band-head at $\lambda7054$ suggests a later than M3 spectral class of the cool component. We attempted to qualitatively reproduce the optical continuum using the sum of radiation of an M5\,III standard star adopted from \cite{Pickles1998} and a $T_e=15000$K nebular continuum (the emission from the hot component with $T_{hot} > 100000$K is negligible at these wavelengths). As one can see in the figure, the modelled SED satisfactorily reproduces the spectrum of 2MASS J21012803+4555377. A similar coincidence can be achieved with a cooler red giant. In this case, $T_{hot}$ derived from EW(He\,II) changes depending on the input from the cool component to the continuum at $\lambda4686$ and increases to $\sim 140000$K when the cool component provides 50\% of flux.       

\section*{Conclusions}

The spectrum of 2MASS J21012803+4555377 demonstrates all the features typical of a SySt, namely, the TiO absorption bands, strong emission lines (H\,I, [O\,III], [Ne\,III], He\,II, etc.), the prominent Raman-scattered O\,VI $\lambda6825$ line. A considerable excess of IR emission with maximum at $\lambda \approx2$~$\mu$m (see fig.~\ref{fig:sed}) allows us to consider this star a D-type SySt based on classification suggested by \cite{Webster1975}, later used in \cite{Belczynski2000} and specified by \cite{Catalog2019}.

We estimated the physical parameters of the components of the binary based on the spectrum obtained near maximum light: the spectral class of the cool component is later than M\,III; with the adopted distance of 2.8~kpc, the luminosity of the sources located inside the dust shell is $\sim 2130$~L$_\odot$; $T_{hot}\geq110000$K derived from the He\,II $\lambda4686$ line.

\section*{Acknowledgments}

N.~Maslennikova thanks the support from the Theoretical Physics and Mathematics Advancement Foundation ''BASIS'' (grant 22-2-10-21-1). Scientific equipment used in this study was bought partially through the M.~V.~Lomonosov Moscow State University Program of Development.

This research has made use of the SIMBAD database (CDS, Strasbourg, France) and Astrophysics Data System (NASA, USA). This work has made use of the data obtained by the European Space Agency (ESA) mission {\it Gaia} (https://www.cosmos.esa.int/gaia), processed by the {\it Gaia} Data Processing and Analysis Consortium (DPAC, https://www.cosmos.esa.int/web/gaia/dpac/consortium). Funding for the DPAC has been provided by national institutions, in particular the institutions participating in the {\it Gaia} Multilateral Agreement.

This publication makes use of data products from the Wide-field Infrared Survey Explorer, which is a joint project of the University of California, Los Angeles, and the Jet Propulsion Laboratory/California Institute of Technology, funded by the National Aeronautics and Space Administration.

This publication makes use of data products from the Two Micron All Sky Survey, which is a joint project of the University of Massachusetts and the Infrared Processing and Analysis Center/California Institute of Technology, funded by the National Aeronautics and Space Administration and the National Science Foundation.

This research is based on observations with AKARI, a JAXA project with the participation of ESA.

\end{document}